\documentclass[12pt]{article}

\advance\voffset by -2.0cm \advance\hoffset by -1.25cm
\usepackage{amsmath}
\usepackage{amssymb}
\usepackage{amsthm}
\usepackage{amsfonts}

\theoremstyle{definition}

\newcommand{\CC}{\mathbb{C}} 
\newcommand{\RR}{\mathbb{R}} 
\newcommand{\ZZ}{\mathbb{Z}} 

\newcommand{\be}{\begin{equation}}
\newcommand{\ee}{\end{equation}}

\newlength{\oldcolsep}

\numberwithin{equation}{section}

\begin{document}

\title{Classification of Commutator Algebras \\
Leading to the New Type of \\ Closed Baker-Campbell-Hausdorff Formulas}
\author{Marco Matone}\date{}

\maketitle

\begin{center} Dipartimento di Fisica e Astronomia ``G. Galilei'' \\
 Istituto
Nazionale di Fisica Nucleare \\
Universit\`a di Padova, Via Marzolo, 8-35131 Padova,
Italy \\  matone@pd.infn.it \end{center}

\begin{abstract}
\noindent We show that there are  {\it 13 types} of commutator algebras leading to the new closed forms of the Baker-Campbell-Hausdorff (BCH) formula
$$\exp(X)\exp(Y)\exp(Z)=\exp({AX+BZ+CY+DI}) \ ,
$$
derived in arXiv:1502.06589, JHEP {\bf 1505} (2015) 113. This includes, as a particular case, $\exp(X) \exp(Z)$, with $[X,Z]$ containing other elements in addition to $X$ and $Z$.
The algorithm exploits the associativity of the BCH formula and is based on the decomposition
$\exp(X)\exp(Y)\exp(Z)=\exp(X)\exp({\alpha Y}) \exp({(1-\alpha) Y}) \exp(Z)$,
 with $\alpha$ fixed in such a way that it reduces to $\exp({\tilde X})\exp({\tilde Y})$, with $\tilde X$ and $\tilde Y$ satisfying the Van-Brunt and Visser condition $[\tilde X,\tilde Y]=\tilde u\tilde X+\tilde v\tilde Y+\tilde cI$.
It turns out that $e^\alpha$ satisfies, in the generic case, an algebraic equation whose exponents depend on the parameters defining the commutator algebra.  In nine {\it types} of commutator algebras, such an equation leads to rational solutions for $\alpha$.
We find all the equations that characterize the solution of the above decomposition problem
by combining it with the Jacobi identity.

\end{abstract}

\newpage

\section{Introduction}

\noindent
The Baker-Campbell-Hausdorff (BCH) formula expresses
\be
Z = \ln(\exp(X) \exp(Y)) ,
\ee
for noncommutative $X$ and $Y$ in a Lie algebra over the complex numbers $\mathfrak g$,
in terms of an infinite series of commutators of commutators. The first papers on the subject are due to Campbell (1897), Poincar\'e (1899), Baker (1902) and Hausdorff (1906). Other important contributions are due to Pascal and Dynkin.
However, after 120 years, there was still no real progress since the series in question is an infinite one and as such of little use for explicit calculations. The exception essentially concerns the trivial case when $ [ X , Y ] $ commutes both with $X$ and $Y$.

\noindent
Very recently, in \cite{Matone:2015wxa}, it has been shown that a simple algorithm extends, to a much more wider class of important cases, including all semisimple complex Lie algebras \cite{Matone:2015oca},
the remarkable simplification of the Baker-Campbell-Hausdorff (BCH) formula recently observed by Van-Brunt and Visser \cite{Van-Brunt:2015ala} (se also \cite{Van-Brunt:2015bza} for related work). In particular, in \cite{Matone:2015wxa} it has been shown the following result.

\noindent
If $X$, $Y$ and $Z$ are elements of $\mathfrak g$, satisfying the commutation relations
\begin{equation}
[X,Y]=uX+vY+cI \ , \qquad [Y,Z]=wY+zZ+dI \ ,
\label{introuno}\end{equation}
with $I$ a central element, and $c$, $d$, $u$, $v$, $z$ complex numbers, then
\begin{equation}
\exp(X)\exp(Y)\exp(Z)=\exp({AX+BY+CZ+DI}) \ ,
\label{introdue}\end{equation}
where $A$, $B$, $C$ and $D$ are $c$-numbers depending on $c$, $d$, $u$, $v$, $w$ and $z$.

\noindent
As a particular case, it has been shown that
if the vector space over $\CC$, spanned by $X$, $Y$, $Z$ and $I$, is closed under the commutation operation, then

\begin{equation}
\exp(X)\exp(Z)=\exp({A'X+B'Y+C'Z+D'I}) \ ,
\label{introtre}\end{equation}
where the $c$-numbers $A'$, $B'$, $C'$ and $D'$ depend on $c$, $d$, $u$, $v$, $w$ and $z$.

\noindent The first step of the algorithm, which exploits the associativity of the BCH formula, is based on the decomposition \cite{Matone:2015wxa}
\begin{equation}
\exp(X)\exp(Y)\exp(Z)=\exp(X)\exp({\alpha Y})\exp({(1-\alpha)Y})\exp(Z) \ .
\label{lladecomposizione}\end{equation}
Next, one fixes $\alpha$ in such a way that $\tilde X$ and $\tilde Y$, defined by
\begin{equation}
\exp(X)\exp({\alpha Y})=\exp({\tilde X}) \ , \qquad \exp({(1-\alpha) Y}) \exp(Z)=\exp({\tilde Y}) \ ,
\label{lladecomposizionedues}\end{equation}
satisfy the Van-Brunt and Visser
condition
\begin{equation}
[\tilde X,\tilde Y]=\tilde u\tilde X+\tilde v\tilde Y+\tilde cI \ .
\label{lacondizione}\end{equation}
This provides the solution of the BCH problem, since now
\begin{equation}
\exp(X)\exp(Y)\exp(Z)=\exp(\tilde u \tilde X + \tilde v \tilde Y + f(\tilde u,\tilde v)[\tilde X,\tilde Y]) \ ,
\label{lllalla}\end{equation}
where $\alpha$, $\tilde u$ and $\tilde v$ are $c$-numbers depending on the parameters defining the commutators between $X$, $Y$ and $Z$ \cite{Matone:2015wxa} and
\begin{equation}
f(u,v)={(u-v)e^{u+v}-(ue^u-ve^v)\over uv(e^u-e^v)} \ ,
\label{lalorofunzione}\end{equation}
is the Van-Brunt and Visser function \cite{Van-Brunt:2015ala}.

\noindent In this paper we study some of the consequences of the algorithm in \cite{Matone:2015wxa}.
In particular, solving the constraints coming from the Jacobi identity allows the classification of the commutator algebras leading to (\ref{lllalla}).

\noindent In section 2 we review the algorithm introduced in \cite{Matone:2015wxa}. In section 3 we provide the explicit expressions of $\tilde u$, $\tilde v$, $\tilde c$ and
of the equation satisfied by $\alpha$. This leads to the explicit expression of (\ref{lllalla}). It turns out that, in the generic case,
the equation for $\alpha$ is an algebraic one for $x^u$ and $x^z$,
where $x:=e^{\alpha}$. In nine {\it types} of commutator algebras, such an equation leads to rational solutions for $\alpha$.
The last section is devoted to a detailed classification of the types of commutator algebras leading to the closed BCH formula (\ref{lllalla}).

\section{Review of the algorithm for BCH}

Here we shortly review the algorithm leading, in relevant cases, to a new class of closed forms of the Baker-Campbell-Hausdorff (BCH) formula \cite{Matone:2015wxa}.
Let us start with the recent finding by Van-Brunt and Visser \cite{Van-Brunt:2015ala} (se also \cite{Van-Brunt:2015bza} for related results). They found
a remarkable relation that simplifies considerably the BCH formula. Their result is the following. If $X$ and $Y$, elements of a Lie algebra $\mathfrak g$, satisfy the commutation relation
\begin{equation}
[X,Y] = uX+vY+cI \ ,
\label{beoftheform}\end{equation}
then \cite{Van-Brunt:2015ala}
\begin{equation}
\exp(X) \exp(Y)= \exp({X+Y+f(u,v)[X,Y]}) \ ,
\label{bbbvvv}\end{equation}
where $f(u,v)$ is the symmetric function
\be
f(u,v)={(u-v)e^{u+v}-(ue^u-ve^v)\over uv(e^u-e^v)} \ .
\ee
We note that in deriving such a result there is a convergence condition for the series expansion of the logarithm in Eq.(22) of \cite{Van-Brunt:2015ala}, that is
\begin{equation}
|1-e^{v-tu}|<1 \ ,
\label{taylor}\end{equation}
with $t\in[0,1]$. Such a condition can be always satisfied by a rescaling of $X$ and $Y$. In this respect, note that
(\ref{bbbvvv}) is a relation involving only the exponentials of $X$, $Y$, $I$ and their linear combinations.
Since, for any suitable norm of $X$ and $Y$, the power expansion of the exponential converges on all $\CC$, it follows that if (\ref{bbbvvv}) holds in a neighborhood of the identity, then, in general, it
should hold in a wider region with respect to the one related to the expansion of $\ln(\exp(X) \exp(Y))$. Of course, this is related to the possible singularities of the norm of
$X+Y+f(u,v)[X,Y]$.

\noindent
Let $X,Y,Z\in{\mathfrak{g}}$. In \cite{Matone:2015wxa} it has been first considered the
decomposition
\begin{equation}
\exp({X})  \exp({Y})  \exp({Z}) = \exp({X})\exp({\alpha Y}) \exp({\beta Y}) \exp({Z}) \ ,
\label{decomposizione}\end{equation}
where $\alpha+\beta=1$.
For all $a\in\CC$, set
\be
s(a):={\sinh(a/2)\over a/2} \ , \qquad s_\alpha(a):={\sinh(\alpha a/2)\over a/2} \ .
\ee
Observe that if
\begin{equation}
{[X,Y]=uX+vY+cI}  \ , \qquad  {[Y,Z]=wY+zZ+dI} \ ,
\label{riecco}\end{equation}
then, by (\ref{bbbvvv}) and (\ref{decomposizione}),
\be
\exp(X) \exp(\alpha Y)= \exp({\tilde X}) \ , \qquad \exp(\beta Y)\exp(Z)=\exp({\tilde Y}) \ ,
\ee
where
\begin{align}
\tilde X &: = g_{\alpha}(u,v)X+ h_{\alpha}(u,v)Y+ l_{\alpha}(u,v)cI \ , \cr
\tilde Y&: = h_{\beta}(z,w)Y+g_{\beta}(z,w)Z+l_{\beta}(z,w)dI \ ,
\label{colon1}\end{align}
with
\begin{align}
g_\alpha(u,v)&:=1+\alpha uf(\alpha u, v)=e^{{\alpha u\over 2}}{s(v)\over s(v-\alpha u)} \ ,  \cr
h_{\alpha}(u,v)&:=\alpha(1+vf(\alpha u, v))= e^{v\over2} {s_\alpha(u)\over s(v-\alpha u)} \ , \cr
l_{\alpha}(u,v)& :=\alpha f(\alpha u, v)={1\over u}\Big( e^{{\alpha u\over 2}}{s(v)\over s(v-\alpha u)}-1\Big) \ .
\label{lageh}\end{align}
Next, we use the associativity of the BCH formula.  In particular, note that if
\begin{equation}
{[\tilde X,\tilde Y]}=\tilde u \tilde X+\tilde v \tilde Y+\tilde c I \ ,
\label{labella}\end{equation}
then, by (\ref{beoftheform}) and (\ref{bbbvvv}),
\begin{equation}
\exp(X) \exp(Y) \exp(Z)= \exp({\tilde X+\tilde Y+f(\tilde u,\tilde v)[\tilde X,\tilde Y]}) \ ,
\label{llasol}\end{equation}
that solves the BCH problem.
Let us consider the commutator
\begin{equation}
[X,Z]=m X+nY+pZ+e I \ .
\label{richiamare}\end{equation}
Note that it is constrained by the Jacobi identity
\be
[X,[Y,Z]]+[Y,[Z,X]]+[Z,[X,Y]]=0 \ ,
\ee
that implies the following linear system for
$m,n,p$ and $e$
\begin{align}
uw+mz & =0 \ ,  \cr
vm-wp + n(z-u) & = 0 \ ,  \cr
pu+zv & = 0 \ ,  \cr
c(w+m)+e(z-u)-d(p+v)&=0 \ .
 \label{system}\end{align}
Next, in order to determine $\alpha$, $\tilde u$, $\tilde v$ and $\tilde c$,
we  replace $\tilde X$ and $\tilde Y$ on the right hand side of (\ref{labella}) by their expressions (\ref{colon1}), and compare the result
with the direct computation, by (\ref{riecco}) and (\ref{richiamare}), of $[\tilde X,\tilde Y]$.
This leads to a linear system of equations whose compatibility condition fixes the equation for $\alpha$. The latter is the basic equation of the algorithm
\begin{equation}
h_{\alpha}(u,v) [h_{\beta}(z,w)(u+z)+g_{\beta}(z,w)(m-w)]+g_{\alpha}(u,v)
[h_{\beta}(z,w)(p-v)-g_{\beta}(z,w)n]=0 \ .
\label{fonda}\end{equation}
The expressions for $\tilde u$, $\tilde v$ and $\tilde c$ follow by the other relations
\begin{align}
\tilde u&=h_{\beta}(z,w)u+g_{\beta}(z,w)m \ ,  \cr
\tilde v&=g_{\alpha}(u,v)p+h_{\alpha}(u,v)z \ ,   \cr
\tilde c&=(h_\beta (z,w)-g_\beta(z,w)l_\alpha(u,v) m) c +(h_\alpha(u,v)-g_\alpha(u,v) l_\beta(z,w) p)d+g_\alpha(u,v)g_\beta(z,w) e \ . \cr
\label{psistema}\end{align}
Note that setting $Y=\lambda_0Q$ and $\lambda_-:=\lambda_0\alpha$, $\lambda_+:=\lambda_0\beta$, the decomposition (\ref{decomposizione})
includes, as a particular case,
\begin{equation}
\exp(X) \exp(Z)=\lim_{\lambda_0\to 0} \exp(X)\exp({\lambda_- Q}) \exp({\lambda_+ Q}) \exp(Z) \ ,
\label{decomposizionetre}\end{equation}
explicitly showing that the algorithm solves the BCH problem for $\exp(X) \exp(Z)$
in some of the cases when $[X,Z]$ also includes other elements than $X$ and $Z$.

\section{The explicit form of the new closed BCH formulas}

The results in the previous section can be summarized by the following theorem. \\

\noindent
{\bf Theorem.} If $X$, $Y$ and $Z$ belong to a Lie algebra and satisfy the commutation relations
$$
{[X,Y]=uX+vY+cI}  \ , \qquad  {[Y,Z]=wY+zZ+dI} \ ,
$$
$c,d,u,v,w,z\in \CC$, then
$$
\exp(X)\exp(Y)\exp(Z) =
$$
\begin{align}
\exp
 \bigg\{
{1\over s(\tilde v-\tilde u)}
\bigg\{
{e^{\tilde u+\alpha u\over2}s(\tilde v)s(v)\over s(v-\alpha u)}  & X + \qquad\qquad\qquad\qquad {} \cr
+\bigg({e^{\tilde u+v\over2}s(\tilde v) s_\alpha(u)\over s(v-\alpha u)} +
{e^{\tilde v+w\over2}s(\tilde u)s_{\beta}(z)\over s(w-\beta z)}\bigg) & Y +  \cr
+  {e^{\tilde v +\beta z\over 2} s(\tilde u)s(w)\over s(w-\beta z)} & Z +  \cr\
+
\bigg[
{e^{\tilde u\over2}s(\tilde v)\over u} \bigg({e^{\alpha u\over2} s(v)\over
s(v-\alpha u)}-1\bigg) c+
{e^{\tilde v\over2} s(\tilde u)\over z}\bigg({e^{\beta z} s(w)\over s(w-\beta z)}-1\bigg) & d +\cr
 + {1\over \tilde u} \bigg(e^{\tilde u\over2}s(\tilde v)-s(\tilde v-\tilde u)\bigg) \tilde c
\bigg] & I\bigg\}\bigg\}
\label{arfas}\end{align}
with $\alpha=1-\beta$ solution of the equation
\begin{equation}
{ e^{v\over2}s_\alpha(u)\Big[e^{w\over2}s_\beta(z)(u+z)+e^{\beta z\over2}s(w)(m-w)\Big]+
e^{\alpha u\over2}s(v)\Big[e^{w\over2}s_\beta(z)(p-v)-e^{\beta z\over2}s(w)n\Big]\over s(v-\alpha u) s(w-\beta z)}=0
\label{peralfa}\end{equation}
corresponding to Eq.(\ref{fonda}), and
\begin{align}
\tilde u &={e^{w\over2}s_\beta(z)u+e^{\beta z\over2} s(w) m\over s(w-\beta z)} \ , \cr
\tilde v &= {e^{\alpha u\over2}s(v)p+e^{v\over2}s_\alpha(u) z\over s(v-\alpha u)} \ , \cr
\tilde c &= \Big(e-{cm\over u}-{dp\over z}\Big)
{e^{\alpha u\over2}s(v)\over s(v-\alpha u)}{e^{\beta z\over2}s(w)\over s(w-\beta z)} + \cr &
+\Big[\Big({w\over z}+{m\over u}\Big){e^{\beta z\over2}s(w)\over s(w-\beta z)}
+\beta-{w\over z}\Big]c+
\Big[\Big({v\over u} +{p\over z}\Big){e^{\alpha u\over2}s(v)\over s(v-\alpha u)}
+\alpha - {v\over u}\Big]  d \ .
\label{vtildsss}\end{align}
\\
The parameters $e,m,n,p$ fix the commutator
$[X,Z]=m X+nY+pZ+e I$,
and are constrained by the linear system (\ref{system}) coming from the Jacobi identity.
\hfill{$\Box$} \\

\noindent
Set $x:=e^\alpha$.
Note that if
\begin{equation}
v-\alpha u \neq 2k{\pi i} \qquad {\rm and} \qquad w-\beta z \neq 2k\pi i \ ,
\label{complessi}\end{equation}
$k\in\ZZ\backslash\{0\}$, then Eq.(\ref{fonda}) is equivalent to
\begin{align}
&x^{u+z} e^{w-2z\over2}\Big({u+z\over uz} e^{v\over2}+{p-v\over z}s(v)\Big)+ \cr
+&x^{u}\Big(ns(v)s(w)-{u+z\over uz}e^{v+w\over2}-{m-w\over u}s(w)e^{v\over2}-{p-v\over z} s(v) e^{w\over2}\Big)+ \cr
-&x^{z} {u+z\over uz}e^{v+w-2z\over2} + \cr
+&{u+z\over uz}e^{v+w\over2}+{m-w\over u}s(w) e^{v\over2}=0 \ .
\label{eh}\end{align}

\section{Jacobi identity and types of commutator algebras}

In the following we provide the classification of the solutions of the BCH formula associated to the commutator algebras
$$
[X,Y]=uX+vY+cI \ , \qquad [Y,Z]=wY+zZ+dI \ ,
$$
$$
[X,Z]= mX+nY+pZ+eI \ ,
$$
and derive the corresponding explicit expressions of Eq.(\ref{arfas}). This is done by solving the Jacobi identity, that is the linear system (\ref{system}).
The resulting constrained parameters fix the form of the equation for $\alpha$,
Eq.(\ref{peralfa}), and determine, by (\ref{vtildsss}), the values of $\tilde u$, $\tilde v$ and $\tilde c$.  Finally, Eq.(\ref{arfas}) provides the corresponding explicit expression of the closed form of the BCH formula.
In the following, we will also assume the condition (\ref{complessi}).

\noindent There are thirteen types of commutator algebras. \\

\centerline{\it The thirteen cases of the Jacobi identity}
\begin{center}
\begin{tabular}{|c|c|c|c|c|}
 \hline {\it 1} & {\it 2} & {\it 3} & {\it 4} &  {\it 5} \\
 \hline $u=z=0$ & $u=0, \; z\neq0$ & $u\neq0,\; z=0$ & $u=z\neq0$ &  $u\neq z,\; uz\neq0$ \\ \hline   $cw\neq dv$ & $w=0$ & $v=0$ & {} & \\
$cw=dv\neq0$ &  $w\neq0$ &  $v\neq 0$ & {} & \\
$cw=dv=0$ & {} & {} & {} & {}  \\ \hline
\end{tabular}
\end{center}

\vspace{.6cm}

\noindent
The case $u=z=0$ is composed of three {\it types} of commutator algebras. These are the {\it type 1a}, corresponding to $cw\neq dv$, the {\it  type 1b}, corresponding to $cw=dv\neq0$, and
the {\it  type 1c}, corresponding to $cw=dv=0$. In turn, the {\it  type 1c} is composed of five {\it types}, depending on which of the parameters in the pairs $cw$ and $dv$ are vanishing.
To illustrate this,
we write only the non-vanishing components of the vector $(c,d,u,v,w,z)$. For example, $(d)$ corresponds to $(0,d\neq0,0,0,0,0)$ and
$(c,v)$ to $(c\neq0,0,0,v\neq0,0,0)$. There are nine possible solutions of $cw=dv=0$. These are $i: (v,w)$, $ii: (c,d)$, $iii: (d,w)\; {\rm or}\; (d)\; {\rm or}\; (w)$, $iv: (c,v)\; {\rm or}\; (c)\; {\rm or}\; (v)$,
$v: (0,0,0,0,0,0)$.
The cases  $(d)$ and $(w)$ are grouped with $(d,w)$ because the corresponding solution of (\ref{system}) holds also when $w=0$ or $d=0$. Similarly, $(c)$ and
$(v)$ are grouped with $(c,v)$.

\noindent
Let us introduce the following symbol\\
$$\Big[case\Big| Jacobi \; constraints\Big|parameters \; of\;  [X,Z]\; unfixed\Big]_{D}  $$\\
where the first slot specifies under which conditions, e.g., $u=z=0$, $cw\neq dv$, the linear system (\ref{system}) is solved. This classifies the
{\it types} of commutator algebras.
The second slot reports the constraints on the commutator parameters fixed by the Jacobi identity, while
the last one reports which ones, between the parameters $m$, $n$, $p$ and $e$ in  $[X,Z]$, are unfixed by the Jacobi identity. In some cases (\ref{system}) also fixes $v$ in $[X,Y]$ or $w$ in $[Y,Z]$. $D$ is the number
of parameters in the commutators between $X$, $Y$ and $Z$, unfixed by the Jacobi identity.
In such a classification, we report the solution of (\ref{system}) in the above symbol and then give the corresponding explicit expression for $\alpha$, solution of Eq.(\ref{peralfa}), and of
$\tilde u$, $\tilde v$ and $\tilde c$, solutions of Eq.(\ref{vtildsss}). In this respect, it is useful to keep in mind that
$\lim_{a\to 0} s(a) =1$, $\lim_{a\to 0} s_\alpha(a) = \alpha$.

\noindent
The first case corresponds to $u=z=0$, which in turn leads to three different types of commutator algebras.
Setting $u=z=0$ in
Eqs.(\ref{peralfa})(\ref{vtildsss}) yields
\begin{align}
\alpha & ={s(v)\big(ns(w)-e^{w\over2}(p-v)\big)\over e^{v\over2}s(w)(m-w)-e^{w\over2}s(v)(p-v)} \ ,  \cr
\tilde u & = m \ , \cr
\tilde v & = p \ , \cr
\tilde c &= \alpha\Big( {dv-cm\over 1-e^{-v}}+{cm\over v}\Big) + \beta\Big({cw-dp\over 1-e^{-w}}+{dp\over w}\Big)+e \ .
\label{vtildsssI}\end{align}

\noindent
The Jacobi identity (\ref{system}) gives $vm=wp$ and $c(w+m)=d(p+v)$.
In this case $e$ and $n$
are always unfixed.  The are three types of commutator algebras, {\it type 1a}, {\it  type 1b} and {\it  type 1c}. The latter is composed of five subtypes.

\vspace{.2cm}

\noindent \underline{\it Type 1a.}
{$$\Big[u=z=0, \; cw\neq dv \Big| m=-w, \; p=-v\Big|e,n\Big]_6  $$}
$$
[X,Y]=vY+cI  \qquad
[Y,Z]=wY+dI  \qquad
[X,Z]=-wX+nY-vZ+eI
$$
\begin{align}
\alpha & ={s(v)\over2}{2e^{w\over2}v+ns(w)\over e^{w\over2}vs(v)-e^{v\over2}ws(w)} & \cr
\tilde u & = -w \nonumber \cr
\tilde v & = -v \nonumber  \cr
\tilde c & = \Big({\alpha\over  1-e^{-v}}+{\beta\over 1-e^{-w}}\Big)(cw+dv) -\alpha{w\over v} c-\beta{v\over w} d + e \nonumber
\end{align}

\vspace{.2cm}

\noindent\underline{\it Type 1b.}
{$$\Big[u=z=0, \;  cw=dv\neq0 \Big| p={vm\over w}\Big|e,m,n\Big]_6  $$}
{$$[X,Y]=vY+cI  \qquad [Y,Z]=w\big(Y+{c\over v}I\big)  \qquad [X,Z]=mX+ nY+{v\over w}m Z+eI $$}
\begin{align}
\alpha & = {s(v)}{e^{w\over2}v(w-m)+nws(w)\over (m-w)(e^{v\over2}ws(w)-e^{w\over2}vs(v))} \nonumber \cr
\tilde u & = m   \nonumber \cr
\tilde v & = {mv\over w}  \nonumber \cr
\tilde c & = \Big[\Big({\alpha\over  1-e^{-v}}-{\alpha\over v}\Big)(w-m) +\Big({\beta\over  1-e^{-w}}-{\beta\over w}\Big)(v-m)+\Big[1+\alpha\Big({w\over v}-1\Big)\Big]c+e \nonumber
\end{align}
The next type is $u=z=0$, with $cw=dv=0$. By (\ref{system}), this leads to several subtypes.

\vspace{.2cm}

\noindent\underline{\it Type 1c-i.}
{$$\Big[(v,w) \Big| p={mv\over w}\Big|e,m,n\Big]_5  $$}
$$
[X,Y]=vY \qquad [Y,Z]=wY  \qquad [X,Z]=mX+nY+{mv\over w}Z+eI
$$
\begin{align}
\alpha & = {ns(v)ws(w)-e^{w\over2}vs(v)(m-w)\over (m-w)(w-v)s(w-v)} \nonumber \cr
\tilde u & = m \nonumber \cr
\tilde v & = {mv\over w} \nonumber \cr
\tilde c & =  e \nonumber
\end{align}

\vspace{.2cm}

\noindent\underline{\it Type 1c-ii.}
{$$\Big[(c,d) \Big| p={cm\over d}\Big|e,m,n\Big]_5  $$}
$$
[X,Y]=cI  \qquad [Y,Z]=dI  \qquad [X,Z]=mX+nY+{cm\over d} Z+eI
$$
\begin{align}
\alpha &={dn-cm\over m(d-c)} \nonumber \cr
\tilde u & = m \nonumber \cr
\tilde v & = {cm\over d} \nonumber \cr
\tilde c & = {dn\over m}+e \nonumber
\end{align}

\vspace{.2cm}

\noindent\underline{\it Type 1c-iii.}
{$$\Big[(d,w)\; {\rm or}\; (d) \; {\rm or}\; (w) \Big| p=0\Big|e,m,n\Big]_{4\, {\rm or}\, 5}  $$}
$$
[X,Y]=0 \qquad [Y,Z]=wY+dI \qquad
[X,Z]=mX+nY+eI
$$
\begin{align}
\alpha & = {n\over m-w} \nonumber \cr
\tilde u & = m \nonumber \cr
\tilde v & = 0  \nonumber \cr
\tilde c & = {dn\over m-w}+e \nonumber
\end{align}

\vspace{.2cm}

\noindent\underline{\it Type 1c-iv.}
{$$\Big[(c,v)\; {\rm or}\; (c) \; {\rm or}\; (v) \Big| m=0\Big|e,n,p\Big]_{4\, {\rm or}\, 5}  $$}
$$
[X,Y]=vY+cI \qquad
[Y,Z]=0  \qquad
[X,Z]=nY+pZ+eI
$$
\begin{align}
\alpha &=1-{n\over p-v} \nonumber \cr
\tilde u & = 0 \nonumber \cr
\tilde v & = p \nonumber \cr
\tilde c & ={nc\over p-v}+e \nonumber
\end{align}

\noindent\underline{\it Type 1c-v.}
{$$\Big[(0,0,0,0,0,0) \Big| \Big|e,m,n,p\Big]_{4}  $$}
$$
[X,Y]=0 \qquad
[Y,Z]=0  \qquad
[X,Z]=mX+nY+pZ+eI
$$
\begin{align}
\alpha & ={n-p\over m-p} \nonumber \cr
\tilde u & = m \nonumber \cr
\tilde v & = p \nonumber \cr
\tilde c & =e \nonumber
\end{align}

\vspace{.2cm}

\noindent The next type, corresponding to $u=0$, $z\neq0$, splits in two types, $w=0$ and $w\neq0$.

\vspace{.2cm}

\noindent\underline{\it Type 2a.}
{$$\Big[u=w=0, \;  z\neq 0 \Big|m=n=v=0, \;  e={pd\over z}\Big|p\Big]_{4}  $$}
$$
[X,Y]=cI \qquad
[Y,Z]=zZ+dI  \qquad
[X,Z]=p\Big(Z+{d\over z}\Big)I
$$
\begin{align}
\alpha & =-{p\over z} \cr
\tilde u&=0  \nonumber \cr
\tilde v&=0  \nonumber \cr
\tilde c&= \Big(1+{p\over z}\Big) c \nonumber
\end{align}

\vspace{.2cm}

\noindent\underline{\it Type 2b.}
{$$\Big[u=0, \; w\neq0,\; z\neq 0 \Big|m=v=0, \;  p={nz\over w}, \; e={dn\over w}-{cw\over z}\Big|n\Big]_{5}  $$}
$$
[X,Y]=cI  \qquad
[Y,Z]=wY+zZ+dI  \qquad
[X,Z]=nY+{nz\over w}Z+\Big({dn\over w}-{cw\over z}\Big)I
$$
\begin{align}
\alpha &=-{n\over w} \nonumber \cr
\tilde u&=0 \nonumber \cr
\tilde v&= 0 \nonumber \cr
\tilde c&= \Big(1+{n\over w}-{w\over z}\Big)c \nonumber
\end{align}

\vspace{.2cm}

\noindent The third type, corresponding to $u\neq0$, $z=0$, splits in two types: $v=0$ and $v\neq0$.

\vspace{.2cm}

\noindent\underline{\it Type 3a.}
{$$\Big[v=z=0, \;  u\neq 0 \Big|n=p=w=0, \;  e={cm\over u}\Big|m\Big]_{4}  $$}
$$
[X,Y]=uX+cI  \qquad
[Y,Z]=dI  \qquad
[X,Z]=m\Big(X+{c\over u}\Big)I
$$
Eq.(\ref{peralfa}) yields
$$
s_\alpha(u)(\beta u+m)=0
$$
Note that (\ref{complessi}) excludes the solutions $\alpha_k={2k\pi i/u}$, $k\in\ZZ\backslash\{0\}$. Therefore
\begin{align}
\alpha & ={m+u\over u} \nonumber \\
\tilde u & = 0  \nonumber \\
\tilde v & = 0 \nonumber \\
\tilde c & = {m+u\over u}d \nonumber
\end{align}

\vspace{.2cm}

\noindent\underline{\it Type 3b.}
{$$\Big[z=0, \;  u\neq 0, \; v\neq0 \Big|p=w=0, \; m={nu\over v}, \; e={cn\over v}-{dv\over u}\Big|n\Big]_{5}  $$}
$$
[X,Y]=uX+vY+cI  \qquad
[Y,Z]=dI  \qquad
[X,Z]={nu\over v}X+nY+\Big({cn\over v}-{dv\over u}\Big)I
$$
In this case (\ref{peralfa}) yields
$$
e^{\alpha u}=e^v
$$
that, taking into account (\ref{complessi}), fixes $\alpha=v/u$. Therefore
\begin{align}
\alpha &={v\over u} \nonumber \cr
\tilde u & = u-v+{nu\over v} \nonumber \cr
\tilde v & = 0 \nonumber \cr
\tilde c &= \Big(1-{v\over u}+{n\over v}\Big) c \nonumber
\end{align}

\vspace{.2cm}

\noindent\underline{\it Type 4.}
{$$\Big[u=z\neq 0\Big|m=-w, \; p=-v\Big|e, n\Big]_{8}  $$}
$$
[X,Y]=uX+vY+cI  \qquad
[Y,Z]=wY+zZ+dI \qquad
[X,Z]=-wX+nY-vZ+eI
$$
In this case (\ref{eh}) leads to an equation of second degree in $x^u$, where $x:= e^\alpha$,
$$
x^{2u}+x^u\Big({nu\over2}s(v)s(w)e^{u+v-w\over2}-e^u-e^v+e^{u+v}-e^{u+v-w}\Big)+e^{u+v-w}=0 \ ,
$$
that is
\begin{equation}
x_\pm^u = {-b\pm \sqrt{b^2-4e^{u+v-w}}\over2} \ ,
\label{laxuX}\end{equation}
where
$$
b:={nu\over2}s(v)s(w)e^{u+v-w\over2}-e^u-e^v+e^{u+v}-e^{u+v-w}\ .
$$ \\
By (\ref{vtildsss})
\begin{align}
\tilde u & =\beta u-w  \cr
\tilde v & =\alpha u -v \cr
\tilde c & = \Big(e+{cw+dv\over u}\Big){ e^{u\over2}s(v)s(w)\over s(v-\alpha u)s(w-\beta u)}- {cw+dv\over u} +\beta c+\alpha d
\label{poffarre}\end{align}

\vspace{.2cm}

\noindent\underline{\it Type 5.}
{$$\Big[u\neq z, \, uz\neq0 \Big|m=-{uw\over z}, \, n=-vw\Big({1\over u}+{1\over z}\Big), \, p=-{vz\over u}, \, e=-{cw\over z}-{dv\over u}\Big|\Big]_{6}  $$}
$$
[X,Y]=uX+vY+cI  \qquad
[Y,Z]=wY+zZ+dI
$$
$$
[X,Z]=-{uw\over z}X-vw\Big({1\over u}+{1\over z}\Big)Y-{vz\over u}Z -\Big({cw\over z}+{dv\over u}\Big)I
$$
\noindent Eq.(\ref{eh}) yields
$$
(x^u-e^v)(x^z-e^{z-w})=0
$$
By (\ref{complessi}), the possible solutions are
$\alpha={v\over u}$
and
$\alpha=1-{w\over z}$.
On the other hand, by (\ref{vtildsss})
\begin{align}
\tilde u&=(\beta z-w){u\over z}  \nonumber \cr
\tilde v&=(\alpha u-v){z\over u}  \nonumber \cr
\tilde c &=
\Big(\beta z-w\Big){c\over z}
+\Big(\alpha u - v\Big) {d\over u}={\tilde u}{c\over u}+{\tilde v}{d\over z}  \nonumber
\end{align}
Therefore, there are two equivalent types of solutions
\begin{align}
\alpha  &={v\over u} \nonumber \cr
\tilde u&=u-v-{uw\over z}  \nonumber \cr
\tilde v&=0  \nonumber \cr
\tilde c &= \Big(1-{v\over u}-{w\over z}\Big)d \nonumber
\end{align}
and
\begin{align}
\alpha  &=1-{w\over z} \nonumber \cr
\tilde u&=0  \nonumber \cr
\tilde v&=z-w-{vz\over u}  \nonumber \cr
\tilde c &= \Big(1-{v\over u}-{w\over z}\Big)d \nonumber
\end{align}
The above commutator algebras appear in several contexts of mathematics and physics, and include all the finite dimensional semisimple complex Lie algebras \cite{Matone:2015oca}.
As an example, we focus on
$X:=\lambda_{-k} {\cal L}_{-k}$, $Y:=\lambda_0 {\cal L}_0$, $Z:=\lambda_k {\cal L}_k$,
where the ${\cal L}_k$'s are the generators of the Virasoro algebra
\begin{equation}
[{\cal L}_j,{\cal L}_k]=(k-j){\cal L}_{j+k}+{c\over 12}(k^3-k)\delta_{j+k,0}I \ ,
\end{equation}
$j,k\in\ZZ$. Note that ${\rm sl}_2(\RR)$  corresponds to the restriction $j,k=-1,0,1$.  We have
$$
[X,Y]=k\lambda_0 X \ , \quad  [Y,Z]=k\lambda_0 Z  \ , \quad
[X,Z]=\lambda_{-k}\lambda_k\Big[{2k\over \lambda_0} Y+
{c\over 12} (k^3-k)\Big] \ ,
$$
so that $c=d=v=w=0$, and
\begin{align}
u & =z=k\lambda_0 \ , \cr
n & =\lambda_{-k}\lambda_k{2k\over\lambda_0} \ , \cr
e & =\lambda_{-k}\lambda_k{c\over 12}(k^3-k) \ ,
\label{centrrr}\end{align}
where the central charge $c$ in (\ref{centrrr}) should not be confused with the $c=0$ in the commutator $[X,Y]$.  By the Jacobi identity,
$m=p=0$,
and by (\ref{poffarre})
\begin{align}
\tilde u&=\beta k\lambda_0 \ ,  \cr
\tilde v&=\alpha k\lambda_0 \ , \cr
\tilde c&=eg_\alpha(k\lambda_0,0)g_{\beta}(k\lambda_0,0)=\lambda_{-k}\lambda_k{\alpha k\lambda_0\over 1-e^{-\alpha k\lambda_0}}{\beta k\lambda_0\over 1-e^{-\beta k\lambda_0}}{c\over 12}(k^3-k) \ .
\label{subito}\end{align}
Setting $\tilde u=\beta u$ and $\tilde v=\alpha u$ in Eq.(\ref{arfas}) yields
\begin{align}
\exp(X)\exp(Y)\exp(Z)&=\exp\bigg\{{(\alpha -\beta)u\over e^{-\beta u}- e^{-\alpha u}}\bigg[X+\bigg(e^{-{\alpha u\over2}}s_\alpha(u)
+ e^{-{\beta u\over 2}} s_{\beta}(u)\bigg) Y+Z + \cr
&+ {1\over \beta u}\bigg(e^{-{\alpha u\over2}}s(\alpha u)- e^{-{u\over2}}s((\alpha-\beta)u)\bigg) \tilde c I\bigg] \bigg\}\ .
\label{diuncertopregio}\end{align}
Set $\lambda_-:=\lambda_0\alpha$, $\lambda_+=: \lambda_0\beta$.
By (\ref{diuncertopregio})
$$  \exp({\lambda_{-k}{\cal L}_{-k}})\exp({\lambda_0{\cal L}_0})
\exp({\lambda_k{\cal L}_k})= $$
\begin{align}
&\exp\bigg\{{\lambda_+-\lambda_-\over e^{-k\lambda_-}-e^{-k\lambda_+}}\bigg[k\lambda_{-k}{\cal L}_{-k}+\bigg(2-e^{-k\lambda_+}-e^{-k\lambda_-}\Big){\cal L}_0+k\lambda_k {\cal L}_k+c_kI\Big]\bigg\} \ ,
\label{viralasoluzionee}\end{align}
where by (\ref{laxuX})
\begin{equation}
e^{-k\lambda_{\pm}}={1+e^{-k\lambda_0}-k^2\lambda_{-k}\lambda_k\pm \sqrt{(1+e^{-k\lambda_0}-k^2\lambda_{-k}\lambda_k)^2-4e^{-k\lambda_0}}\over2} \ ,
\end{equation}
that, together with (\ref{subito}), gives
\begin{equation}
c_k={\lambda_{-k}\lambda_k\over \lambda_+-\lambda_-} \bigg({\lambda_+\over 1-e^{-k\lambda_+}}-{\lambda_-\over 1-e^{-k\lambda_-}}\bigg){c\over12}(k^4-k^2) \ .
\end{equation}
Finally, we report the case $\lambda_0=0$ \\
\begin{equation}
\exp({\lambda_{-k}{\cal L}_{-k}})
\exp({\lambda_k{\cal L}_k})=\exp\bigg[ {\lambda_+\over \sinh (k\lambda_+)}(k\lambda_{-k}{\cal L}_{-k}+k^2\lambda_{-k}\lambda_k{\cal L}_0+k\lambda_k {\cal L}_k+c_kI)\bigg]\ ,
\label{viralasoluzioneezeroBBB}\end{equation}
where now
\\
\begin{equation}
c_k=\lambda_{-k}\lambda_k{c\over24}(k^4-k^2) \ .
\end{equation}

\section*{Acknowledgements} It is a pleasure to thank  Andrea Bonfiglioli, Pieralberto Marchetti,  Leonardo Pagani, Paolo Pasti,
 Dmitri Sorokin and Roberto Volpato for interesting
discussions.


\newpage

\begin{thebibliography}{99}



\bibitem{Matone:2015wxa}
 M.~Matone,
``An algorithm for the Baker-Campbell-Hausdorff formula,''
  JHEP {\bf 1505} (2015) 113,
 arXiv:1502.06589.


\bibitem{Matone:2015oca}
  M.~Matone,
  ``Closed Form of the Baker-Campbell-Hausdorff Formula for Semisimple Complex Lie Algebras,''
  arXiv:1504.05174. 

\bibitem{Van-Brunt:2015ala}
A.~Van-Brunt and M.~Visser,
  ``Special-case closed form of the Baker-Campbell-Hausdorff formula,''
   J.\ Phys.\ A {\bf 48} (2015)   225207,
  arXiv:1501.02506.

\bibitem{Van-Brunt:2015bza}
  A.~Van-Brunt and M.~Visser,
 ``Simplifying the Reinsch algorithm for the Baker-Campbell-Hausdorff series,''
  arXiv:1501.05034. 



\end{thebibliography}
\end{document}